# Auditory steady-state response and gamma oscillations in an excitatory-inhibitory balanced neuronal network

Duoyu Feng[1], Jiajia Li [1, 2,*]

[1)] Department of Artificial Intelligence and Automation, College of Information and Control Engineering, Xi'an University of Architecture and Technology, Shaanxi, Xi'an, 710055, China;

[2)] Department of Neurosurgery, PLA General Hospital of Central Theater Command, Wuhan, 430070, China

[*] Correspondence should be addressed to lijiajia_dynamics@xauat.edu.cn

**Abstract:**

This study introduces a novel auditory neuronal network model that integrates speech signal input, cochlear processing, and a cortical excitatory-inhibitory (E-I) balanced network. Our findings reveal that increasing noise intensity attenuates the auditory steady-state responses in gamma oscillations, a mechanism validated by public EEG data. Moreover, enhancing the brain's E-I balance significantly improves auditory attention during speech recognition. This work not only elucidates the neural basis of selective attention in noisy environments but also offers a promising therapeutic strategy for auditory attention disorders, marking a significant advancement in the field of computational neuroscience and auditory processing.

**Highlighted:**

Gamma neural oscillation amplitude encoding may underpin attention mechanisms.

Boosting the model excitatory-inhibitory balance enhances the attention responses in gamma oscillation, suggesting a novel therapeutic approach for attention-deficit disorders.

## 1 Introduction:

Hearing is one of the main channels through which external information can be transmitted to the brain. The auditory speech perception system in the brain is capable of extracting effective components of target speech information even in noisy environments [1,2]. However, there remains an unsolved issue in this field that how our brain cerebral cortex perceives speech information amidst the environmental noise.

In fact, external speech signals are often accompanied by environmental noise (such as traffic noise, machinery noise, etc.) and non-target speech signals. These environmental noises can be categorized into steady-state noise (such as that noise from computers and refrigerators) and variable noise (like traffic noise). Excessive noise can weaken the brain's ability to perceive target speech signals (speech-in-noise, SIN), resulting in an inability to accurately and precisely perceive these signals, which is known as the classic

"cocktail party problem"[3]. According to the proposal of recent literature, to perceive speech in the presence of noise, the brain must construct the selective attention to the target signal [4]. Over the past two decades, there has been an abundance of research on the auditory pathway of speech recognition in the human brain [5-9]. The theory has also been widely applied in the field of human-computer interaction, HCI, aimed at optimizing the speech recognition and interaction functions of humanoid robots [10,11]. That is, a deeper understanding of the speech recognition mechanisms in the human brain is key to theoretically enhancing the speech recognition systems in HCI [12].

In fact, during the natural language processing by HCI speech recognition system, the attention modules facilitate the implementation of neural network separation function [13]. This not only helps improve performance in tasks such as machine translation, semantic understanding, and dialogue generation, but also enables the system to exhibit greater robustness and adaptability when faced with diverse and ambiguous language inputs [14]. Similarly, the attention mechanism in human brain is also beneficial for language understanding processes during communication, and aids in optimizing the neural networks responsible for natural language understanding [15,16], implying that attention construction plays a critical role in sustaining our cerebral cortex to perceive the speech information in the auditory inputs [17-19].

Then, we expect to ask how the auditory attentional module helps the bran to decode the influx of speech signals, and what has contributed to the critical function of attention. Brain imaging techniques have confirmed the neural mechanisms of auditory attention regulation by the finding that the banded region of the lateral temporal auditory cortex exhibits strong activation following attention triggered by speech signal inputs [20]. The specific selective attention modulation was also found to occur in the superior temporal gyrus and the superior temporal sulcus, in response to continuous speech stimuli [21]. Therefore, the attention was shown to coordinates different regions in the cerebral cortex to decode external speech inputs [22]. This proposal inspired various studies to detect the dynamical changes in cortical electroencephalography (EEG) to investigate attentional mechanisms [23-27]. Liebherr et al. leveraged the dynamical feature of EEG to unravel that environmental and specific task demands can affect attention processing, which was represented in the biomarkers, the event-related potential components of mismatch negativity and P3. Wei et al. discovered through EEG techniques that higher working memory load can impair the selective attention by affecting top-down cognitive control, which is symbolled by the feature of a larger N2 response in the cerebral cortex. Some other scientists proposed that the specific rhythm oscillations seem to involve more robust and noise-ratio features for unraveling attentional mechanisms. For example, the frequency feature of cortical EEG theta waves (4-8 Hz) was found to be generally associated with focused attention and task preparation, while gamma waves (30-100 Hz) were proved to be linked to higher cognitive processing and the synchrony of neural networks [28]. In summary, attention modulation can alter electrical activity in specific cortical areas, leading to changes in local and global electrical activity patterns [29,30]. Specifically in the view of cortical gamma oscillations response, experimental research including ours have given direct evidence for the positive correlation between gamma oscillations and the formation of attention [31,32]. Then a new question arises that how this gamma oscillation-related attention switch between on/off state by the modulations of the intrinsic neurons and synapses.

For better understanding the specific effects of different neuron and synapses on the auditor attention -related gamma oscillation, we utilized the excitatory/inhibitory neuron population -coupled models [33-34] to mimic the mean filed response of cortical auditory input. Meanwhile, inhibitory interneurons and the GABA synapses were considered to mimic the inhibitory modulation loop [34], as were the pyramidal neurons and excitatory synapses to mimic the excitatory modulation loop [35]. Moreover, the imbalanced excitatory/inhibitory inputs [3] were also considered to mimic the actual structure topology for excitatory/inhibitory neurons.

At last, this paper also considered the cochlea's frequency resolution function for the speech signal, with which this paper constructs a mathematical model of the speech input–cochlea–cortical population neural network to investigate the attentional -related gamma oscillation switching processes under different background noises, and the potential positive modulation effect of excitatory /inhibitor balance on the attentional states.

## 2 Materials and Methods

To explain the neural mechanisms of speech processing in the brain under noise conditions, we propose the cochlear-thalamic-cortical neural network model, as illustrated in Fig. 1. First, external speech signals enter through the cochlea, traveling along the basilar membrane. The basilar membrane has a frequency-selective function, dispersing the various frequency components of the speech signal to various locations（regions）, where they are then perceived by hair cells. These signals are further converted into electrical signals and transmitted to the thalamus. The thalamus, serving as a higher auditory center, can perform a comprehensive analysis of the frequency feature of the speech signals, and transmits the processed signals to the primary auditory cortex. This enables the auditory cortex to receive stimulus and respond with brain electrical activities, encoding the speech signals for perception, thereby completing the brain's process of speech perception coding. The processes outlined above have been explained and described by the mathematical equations below in details.

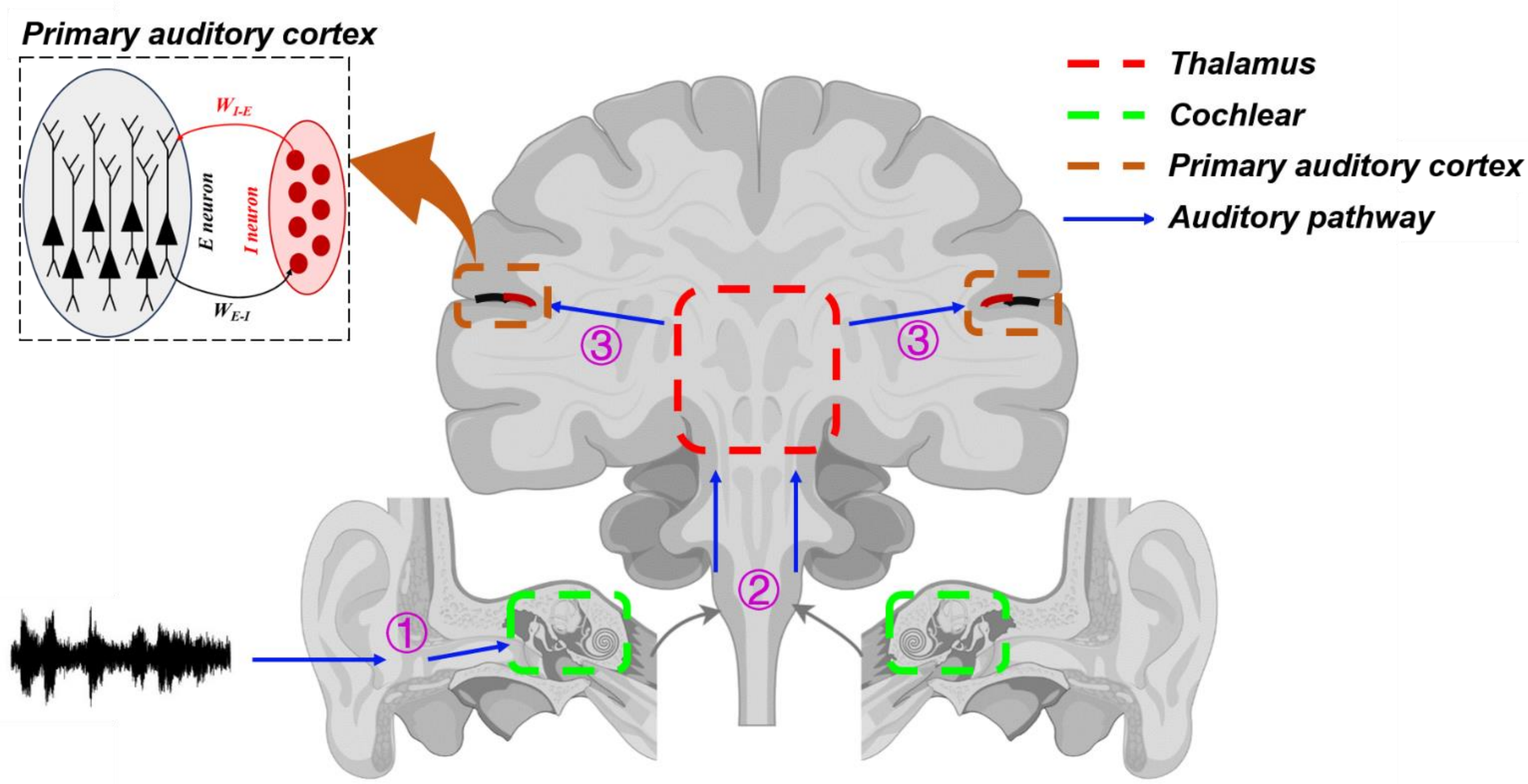

Fig.1 schematic diagram of the auditory pathway (drawn by *Figdraw*). The auditory pathway mainly consists of the cochlea, thalamus, and auditory cortex. Speech signals enter the human ear and are processed by the cochlea, where they are converted into electrical signals that are transmitted to the thalamus. The thalamus then analyzes features such as the time-frequency features of the signals, which are subsequently transmitted to the auditory cortex. In this study, the auditory cortex（cortical）neural network model including excitatory and inhibitory neurons was proposed to investigate the mechanisms of speech signal processing within the auditory cortex.

### 2.1 The speech input–cochlea coupled system

The electrical activities in the auditory cortical neural network arise from speech input signals processed through the cochlea and thalamus. This paper first preprocesses the speech signals through time-frequency transformation and envelope processing. The processed signals are equivalent to the signal processing functions of the cochlea-thalamus system, serving as input for the neurons of primary auditory cortex. The specific preprocessing method is described by the following equation:

$$X = FFT\ transform\ (Voice\ Signal) \tag{1}$$

$$x = envelope(X) \tag{2}$$

First, a Fast Fourier Transform (FFT) is applied to the original speech signal to simulate the frequency separation function of the cochlea, resulting in the time-frequency features $X$ of the original speech signal. Next, the high-power regions in $X$ are enveloped to obtain the characteristic envelope $x$ .

$$x' = \frac{x - x_{\min}}{x_{\max} - x_{\min}} \tag{3}$$

$$I_{thalamus.i}^{Ed} = A_i^E x' \tag{4}$$

$$I_{thalamus.j}^{I} = A_j^I x' \tag{5}$$

The obtained speech characteristic envelope cannot be directly used as input of thalamic stimulation current for the auditory cortical neural network. To accurately quantify the parameters of the speech input signal, normalization is performed on $x$, as shown in Eq. (3), resulting in the normalized characteristic envelope $x'$. Finally, the coefficients $A_i^E$ and $A_j^I$ are used to precisely control the parameters of the input speech signal, indicating that $I_{thalamus.i}^{Ed}$ and $I_{thalamus.j}^{I}$ represent the equivalent input signals for the auditory cortex（cortical）neural network from the cochlea-thalamus system at this stage. Here, $A_i^E : A_i^I = 5:2$ is based on the biophysical parameters of cortical networks in the previous work [37].

## 2.2 The auditory cortical neural network model

In this model, the frequency separation and signal analysis functions in the cochlea-thalamus systems are equivalently explained as the signal preprocessing component in the Eqs. (1)-(5). The auditory cortex （cortical） neural network model is constructed as an excitatory-inhibitory (*E-I*) balanced neural network composed of 200 excitatory pyramidal neurons and 50 inhibitory interneurons. As shown in Fig.1 top left, all the pyramidal neurons form a cluster of excitatory pyramidal neurons (*E*), while all the inhibitory interneurons form a cluster of interneurons (*I*). The layers are connected by using mean-field connections, where the connection weights are determined by the mean firing rates of the neural populations.

Additionally, since the *E* and *I* neuron populations employ a fully connected structure within themselves, there are also excitatory and inhibitory effects of synapses on each other within each cluster of neurons. The excitatory inputs and inhibitory inputs received by the two neural populations counterbalance each other, resulting in a dynamic balance of excitatory and inhibitory current transmission within the network. Compared to single neurons, the *E-I* balanced neural network can track external signals more swiftly and respond quickly to external stimuli. It is also capable of simulating gamma oscillations during perceptual tasks, which aligns more closely with brain characteristics.

In the cluster of *E* neurons, the excitatory pyramidal neurons utilize the two-compartment model proposed by Wang et al. [38], which is characterized by that all currents transmitted to the pyramidal neurons firstly enter through the dendritic part before reach the soma. This model not only better reflects the morphological characteristics of neuron structures but also generates richer neuron firing patterns. The membrane potential of each neuron is described by the following equation:

$$C_{m,E}\frac{dV_{E,i}}{dt}=f^{E}(V_{E,i},m_i,n_i,h_i)+\frac{g_c}{p}\left(V_{Ed,i}-V_{E,i}\right) \tag{6}$$

$$C_{m,E}\frac{dV_{Ed,i}}{dt}=f^{Ed}(V_{Ed},s,\left[Ca^{2+}\right]_n)+\frac{g_c}{(1-p)}\left(V_{E,i}-V_{Ed,i}\right)+I^{Ed}_{syn,i}+I^{Ed}_{thalamus,i} \tag{7}$$

Where $C_{m,E}$ represents the membrane capacitance of the neuron, the action potentials in the two-compartment neuron model are represented by $V_{E,i}$ for the soma and $V_{Ed,i}$ for the dendrites. The soma of each neuron is composed of leakage current, ionic current ( $f^{E}(V_{E,i},m_i,n_i,h_i)$ )described by the equations of **Appendix** in detail, and the corresponding dendritic coupling current ( $\frac{g_c}{p}\left(V_{Ed,i}-V_{E,i}\right)$ ). Each neuron's dendrite consists of leakage current, multiple ionic currents ( $f^{Ed}(V_{Ed},s,\left[Ca^{2+}\right]_n)$ )described by the equations of **Appendix** in detail, synaptic currents $I^{Ed}_{syn,i}$ , thalamic equivalent current $I^{Ed}_{thalamus,i}$ , and the corresponding dendritic coupling current ( $\frac{g_c}{(1-p)}\left(V_{E,i}-V_{Ed,i}\right)$ ). The units of $\frac{g_c}{p}\left(V_{Ed,i}-V_{E,i}\right)$ and $\frac{g_c}{(1-p)}\left(V_{E,i}-V_{Ed,i}\right)$ are μA/cm², where $g_c$ represents the coupling conductance and $p$ denotes the ratio of the soma surface area to the total surface area, indicating the difference in current intensity due to the varying surface areas of the cell membranes. The corresponding parameters for the pyramidal cell are presented in Table 1.

In the cluster of $I$ neurons, the inhibitory interneurons utilize the fast-spiking interneuron model established by Wang et al. [33]. Theoretical research indicates that when appropriate conditions for the dynamical processes of synaptic transmission are met, these GABAergic interconnections can synchronize the networks of neurons with gamma oscillations. We employ the interneuron model to provide gamma frequency oscillations through synaptic transmission mediated by $GABA_A$ receptors, which synchronize the discharge of spatially distributed neurons. The membrane potential of each neuron is described by the following equation:

$$C_{m,I}\frac{dV_{I,j}}{dt}=f^{I}(V_{I,j},m_j,n_j,h_j)+I^{I}_{syn,j}+I^{I}_{thalamus,j} \tag{8}$$

Each neuron is composed of ionic currents ( $f^{I}(V_{I,j},m_j,n_j,h_j)$ , synaptic current $I^{I}_{syn,j}$, and thalamic equivalent current $I^{I}_{thalamus,j}$. The corresponding parameters for the interneuron are presented in Table 2.

The detailed expression for the synaptic current of pyramidal neurons is:

$$I^{Ed}_{syn,i}=g_I\cdot\left(\sum_{k=1}^{N_I}(w_k^{I\to E}y_{GABA,k})\right)\cdot(V_{GABA}-V_{Ed,i})+g_E\cdot\sum_{k=1,k\neq i}^{N_E}y_{AMPA,k}/N_E\cdot(V_{AMPA}-V_{Ed,i}) \tag{9}$$

The term $g_I\cdot\left(\sum_{k=1}^{N_I}(w_k^{I\to E}y_{GABA,k})\right)\cdot(V_{GABA}-V_{Ed,i})$ represents all the inhibitory currents received by the pyramidal neuron and $w_k^{I\to E}$ denotes the mean field coupling weight, which is equal to the firing rate of the neurons in interneuron populations. $g_I$ denotes the inhibitory synaptic conductance of GABA synapses, and $y_{GABA}$ represents the opening and closing degree of GABA channels. The term $g_E \cdot \sum_{k=1,k\neq i}^{N_E} y_{AMPA,k} / N_E \cdot (V_{AMPA} - V_{Ed,i})$ represents the sum of the excitatory currents from all pyramidal neurons except the i^th neuron in a fully connected manner. The term $g_E$ denotes the excitatory synaptic conductance of AMPA synapses, and $y_{AMPA}$ represents the opening and closing degree of AMPA receptor channels.

The activation and deactivation states of the AMPA and GABA receptor channels are described by the following equations:

$$\frac{\mathrm{d}y_{\mathrm{AMPA,i}}}{dt} = 1.1[T](1 - y_{\mathrm{AMPA,i}}) - 0.19 y_{\mathrm{AMPA,i}} \tag{10}$$

Where $[T]$ represents the amount of neurotransmitter released from the presynaptic terminal into the synaptic cleft, described by the following equation:

$$[T] = T_{\max} / (1 + \exp(-(V - V_T)/K_p)) \tag{11}$$

According to the ref. [39], $T_{max}$=1mM， $V_T$=2， $K_p$=5mV.

$$\frac{\mathrm{d}y_{GABA,j}}{dt} = 5(1 - y_{GABA,j}) / (1 + \exp(-V/2)) - 0.18 y_{GABA,j} \tag{12}$$

The detailed expression for the synaptic current of interneurons is:

$$\begin{aligned} I_{syn,j}^{I} &= g_E \cdot \left(\sum_{k=1}^{N_E} (w_k^{E\to I} y_{AMPA,k})\right) \cdot (V_{AMPA} - V_{I,j}) + g_{autapse_in} y_{GABA,j} (V_{GABA} - V_{I,j}) \\ &+ g_I \cdot \sum_{k=1,k\neq j}^{N_I} y_{GABA,k} / N_I \cdot (V_{GABA} - V_{I,j}) \end{aligned} \tag{13}$$

Here, $g_I$ represents the synaptic conductance of GABA synapses, with the value of 0.004mS/cm$^2$ to adapt to the ref. [39], and $y_{GABA}$ represents the opening and closing degree of GABA channels. $g_E$ denotes the excitatory synaptic conductance of AMPA synapses, and $y_{AMPA}$ represents the opening and closing degree of AMPA channels. The term $w_k^{E\to I}$ represents the mean-field coupling weight of excitatory synaptic currents, which is equal to the firing rate of the neurons of pyramidal populations. $g_E \cdot \left(\sum_{k=1}^{N_E} (w_k^{E\to I} y_{AMPA,k})\right) \cdot (V_{AMPA} - V_{I,j})$ represents the total sum of excitatory currents from all pyramidal neurons in a mean-field manner.

The term $g_{autapse_in} y_{GABA,j} (V_{GABA} - V_{I,j})$ indicates the inhibitory auta-synaptic current applied to the $j^{th}$ $I$ neuron with the inhibitory auta-synaptic conductance of 0.1mS/cm$^2$ [40], while the term $g_I \cdot \sum_{k=1,k\neq j}^{N_I} y_{GABA,k} / N_I \cdot (V_{GABA} - V_{I,j})$ represents the total inhibitory currents from all other $I$ neurons. Finally, the reversals potential for both the excitatory AMPA and inhibitory synapses are: $V_{AMPA}$ =0mV, $V_{GABA}$=-75mV according to the ref. [33] and [38], respectively.

### 2.3 Human EEG Dataset for the auditory pure-tone stimuli and the GBO analysis algorithm

In this paper, for validating the modeling proposal on the neuronal responses to auditor stimuli, a publicly available dataset [41] was also recruited, which included EEG data from 13 subjects. Each subject underwent three identical stimulation experiments, each lasting 13 minutes, with the stimulus being 1000 Hz and 500Hz pure tones and the noise stimulus being 76 dB white noise. All sound stimuli lasted for 60 ms. the GBO curves of brain electrical responses under pure tone and noise stimuli were analyzed to be compared with the modeling results.

Furthermore, to quantify the effect of different noise levels on the electrical activity of the auditory cortex （cortical） neural network in terms of gamma oscillation intensity, this paper introduces a filtering method to analyze the neuronal firing series based our previous human experimental work [31]. First, the 30-100 Hz portion of the neuronal firing series is filtered, then the squares of the power is calculated and smoothed, obtaining the gamma oscillation, where the peak of the processed curve represents the gamma oscillation intensity.

The filter used is an infinite impulse response (IIR) digital filter, which achieves excellent frequency selection characteristics with a relatively low order and possesses sufficient robustness to maintain passband characteristics. The IIR digital filter is described by the system function *H(z)* as shown in Eq. (14):

$$H(z) = \frac{\sum_{k=0}^{M} b_k z^{-k}}{1 + \sum_{k=1}^{N} a_k z^{-k}} \quad (14)$$

By transforming the frequency, the cutoff frequencies of the passband and stopband of the analog filter are obtained from the boundary frequencies of the IIR digital filter. This results in the derivation of the system function *H(s)*. The *H(s)* function is then converted into *H(z)* using the impulse invariant method. Based on the technical specifications of the IIR digital filter, the system function *H(z)* of the IIR digital filter is determined. The quadratic power of the 30-100 Hz range of the filtered discharge signal by the bandpass IIR digital filter is calculated to obtain the power value for this portion, and the resulting curve is smoothed. The result is the dynamical power curve of gamma oscillation (Gamma Band Oscillation, GBO), which is used in this paper as an indicator to analyze the intensity of brain attention represented by varying GBO responses under different stimuli.

## 3 Results：

### 3.1 Temporal-frequency responses of cortical neural networks to the speech signal inputs

Firstly, in order to explore the effect of environmental noise on the electrical activity of the auditory cortex(cortical) neural network, we utilized an equivalent model of cochlear -thalamic system to discuss the features of speech input in terms of cochlear time-frequency features and equivalent thalamic input signal patterns under both noisy and noise-free conditions. As shown in Fig. 2A, it depicts the diagram of the cortical neural network including the *E/I* neurons, their connections and the input (speech signals)/outputs (local field potential, *LFP*).

As illustrated in Fig. 2B, human continuous speech signals with/without noise were selected firstly. After applying Fourier transformation, the frequency domain features of the continuous speech input signal were transformed by the cochlear system, where red areas representing high power were observed distributing in both the low-frequency and high-frequency regions. However, when analyzing a 60 dB noise-mixed speech signal, the frequency domain feature processed by the cochlear system displayed that the high power mainly concentrated in the low-frequency region, while the power of high-frequency region diminishes. After obtaining the feature envelope based on frequency feature of red region to mimic the thalamic input, it became evident that the envelope curve of the speech signal under noise conditions had lost most of the time-frequency features, with an overall amplitude lower than that of the noise-free original speech signal. The results also indicate that at a certain intensity of noise, different frequency regions within the sound experience varying degrees of masking. This also reflects the phenomenon of masking in speech perception within noisy environments [42].

Furthermore, after applying the equivalent thalamic inputs of the envelop curves to the excitatory-inhibitory (E-I) balanced neural network model of the auditory cortex, we compared the neural network discharge response patterns in the two conditions characterized by the presence or absence of noise in the continuous speech signal, as shown in Fig. 2C-D.

Under the noise-free condition, the local field potential (LFP) of the auditory cortex exhibits a higher frequency during the speech stimulus period (Fig. 2C upper panel), and the spiking pattern of the neuron populations corresponding to the stimulus period in the map (Fig. 2C lower panel) is denser. Conversely, under the noisy condition, the LFP shows a lower frequency during the speech stimulus period (Fig. 2C upper panel), and spiking pattern of the neuron populations map (Fig. 2C lower panel) is sparser. The comparison between the two conditions indicates that the activation level of the auditory cortex is greater in the noise-free condition, resulting in a stronger response of neuron populations. At this time, the human brain's perception of the speech signal is stronger, allowing individuals to clearly hear external speech. In contrast, the activation level of the auditory cortex is lower in the noisy condition, resulting in a weaker response compared to the noise-free condition. Consequently, the human brain's perception of the speech signal is diminished, making it difficult to hear external speech in a noisy environment.

Further quantitative analysis of the discharge response patterns of the auditory cortical neural network under noisy and noise-free conditions is presented in Fig. 2D. By analyzing the post-stimuli time histogram (PSTH) and gamma oscillations of the neuron population discharge of the auditory cortex, we found that the discharge frequency was consistently greater than 25 Hz under the noise-free condition (Fig. 2D upper panel), but with specific moments of frequencies greater than 50 Hz ( the gamma oscillation frequency band is 30-100 Hz) during the speech singal input. The power amplitude of the Gamma Band Oscillation (GBO) (Fig. 2D lower panel) also follows a trend similar to that of the PSTH: a higher GBO power amplitude corresponds to a higher frequency in the histograms (Fig. 2D upper panel).

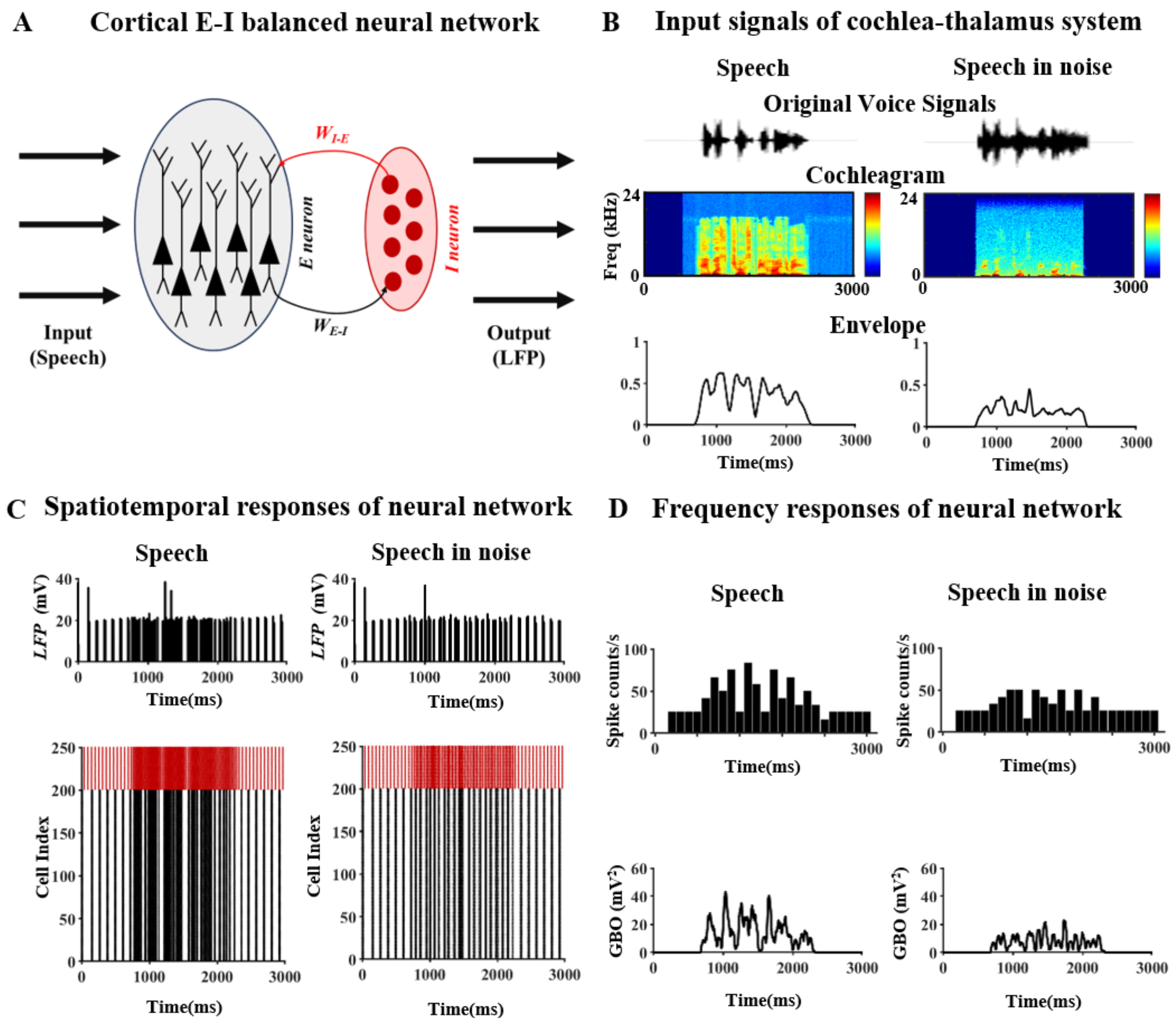

Fig 2 Responses of the E-I balanced neural network under noise-free and noise-mixed continuous speech inputs. A. Schematic diagram of the cortical E-I balanced neural network; B. Schematic diagram of the speech signal processing process: the original input of the speech signal in the top panel, the transformed frequency-time power patterns of the speech signal processed by cochlea in the middle panel, and the equivalent thalamic input that is directly imposed to the cortex is the envelope of the high-power regions in the frequency-time features from the middle panel; C. the temporal responses of both the Local-field potential (LFP) and spatiotemporal spiking patterns of the neuron populations in the auditory cortex in response to noise-free and noise-mixed continuous speech inputs; D. the frequency responses of both the PSTH and the dynamical gamma oscillation response of the neuron population in the auditory cortex to noise-free and noise-mixed continuous speech inputs.

Those specific moments of GBO peaking and timing information represents a coding mechanism of brain auditory system for the external contineous speech signal input. Under noisy conditions, the discharge frequency of the auditory cortex responses remains persistently low, below 35 Hz (Fig. 2D upper panel), with the corresponding GBO amplitudes (Fig. 2D lower panel) also being low. This comparison indicates that under noise-free conditions, Gamma oscillations are more intense, enabling the brain's attention to focus on the speech signal, resulting in directed perception of external speech. As demonstrated below Fig. 2D lower panel, the cortical GBO peak pattern can encode and identify the speech input, while under noisy conditions, gamma oscillations are attenuated, and the discharge frequency of auditory cortex's response falls below 35 Hz, entering the beta frequency band (12-30 Hz). At this point, the brain can barely construct directed attention, leading to weaker perception capacity for external speech. Noise masks most of the mid- and high-frequency regions in the speech signal, with a minor portion of the low-frequency region also being masked. This prevents gamma oscillations from encoding most features of the speech, resulting in a reduced ability of human auditory system to perceive and accurately recognize external speech. This indicates that environmental noise interferes with the brain's capacity to perceive speech signals. According to the literature [43], environmental noise negatively influences speech perception, specifically underliend by a significant decline in subjects' attention as noise levels increase. The comparasion results illustrate the masking effect of noise on speech signals, leading to a decreased speech perception ability in the brain. The previous work [44] also mentions that under noisy conditions, the cognitive performance accuracy rate of the respondents declines, supporting the above couclusion in this paper.

In order to study the discharge dynamic features of the auditory cortical neural network, and explain the corresponding discharge patterns of cortical responses under both noisy and noise-free conditions, a system bifurcation analysis method was used, and the results are shown in Fig. 3.

As shown in the first column of Fig.3A-B, the system's stable equilibrium point exhibits a Hopf bifurcation for both the excitatory and inhibitory neurons, and when the bifurcation parameter（the assemble currents that include both the synaptic currents and other external current，$I_{assemble}$) is located in the ranges between the two bifurcation points, the postsynaptic neurons transits into firing state. While, the stable equilibrium point becomes unstable as the bifurcation parameter goes out of that range.

In a dynamical view of the bifurcation characteristics, the bifurcation parameters of each the *E/I* neuron are dominant by the synaptic currents imposed to each neuron. Therefore, as shown in the second and third column of Fig. 3 A and B, the excitatory postsynaptic current (EPSC) and the inhibitory postsynaptic current (IPSC) for single *E/I* are depicted. It can be observed that the EPSCs for both *E* and *I* neuron goes through a significant decrease as the input is transited from speech to the case of speech with noise. However, this significant decrease was absent in the IPSC column. This implies that the alterations in spiking patterns and GBO dynamics that are affected by the noise masking effect (Fig. 3) arise mainly from the EPSC dynamics, which could modulate the bifurcation parameter for the *E/I* neurons in a dynamical view.

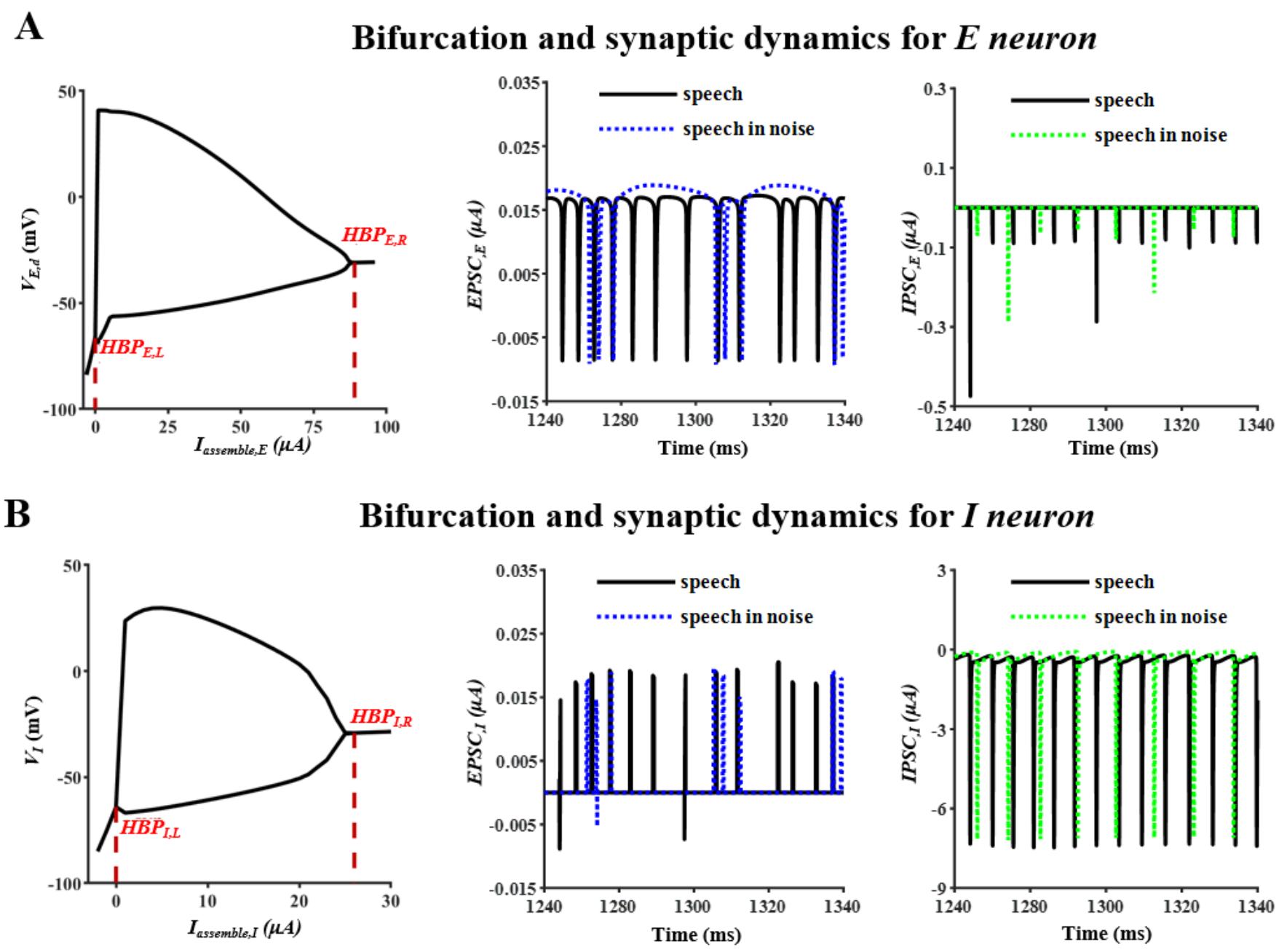

Fig. 3. Neuronal bifurcation and snaptic dynamics for excitatory and inhibitory neurons. A. Bifurcation dynamics and the dynamic changes in the *EPSC* and *IPSC* for sinlge *E neuron*, where $HBP_{E,L}$ represents the critical point of the system transitioning from a stable state to an unstable state, while $HBP_{E,R}$ represents the critical point of the system transitioning from an unstable state to a stable state; B. Bifurcation dynamics and the dynamic changes in the *EPSC* and *IPSC* for sinlge *I neuron*, where $HBP_{I,L}$ represents the critical point of the system transitioning from a stable state to an unstable state, while $HBP_{I,R}$ represents the critical point of the system transitioning from an unstable state to a stable state.

### 3.2 Coding dynamics of GBO to the speech inputs and the noise masking effect

To quantitatively study the effect of noise on the auditory cortex's perception process of continuous speech, time-frequency transformations and the gamma oscillations of the neuron populations were performed on the noise-mixed speech signals in the range of 20-70 dB (increasing by 10 dB). Gamma oscillation was obtained based on the algorithms in Materials and Methods 2.3 to quantify the firing frequency features of the cortical neural network response. The above analysis was designed to unveil the different degrees of masking effects of noise at various sound pressure levels (SPL) on speech signals and the brain's perception of these signals. Fig. 4A shows the time-frequency features of the 20-70 dB noise (increasing by 10 dB) mixed speech signals and their corresponding cortical response GBO curves. It can be observed that as the SPL of the mixed noise increases, the features of the speech signal are progressively masked, and the corresponding GBO curve amplitude gradually decreases. When the SPL is 20-30 dB, only a small portion of high-frequency features of the speech signal is masked, leading to minimal variation in the GBO amplitude of the auditory cortex response. When the SPL is 40-50 dB, a significant portion of mid- to high-frequency features of the speech signal are masked, resulting in a decrease in the GBO amplitude of the auditory cortex response, and the GBO peak in the figure is visibly reduced to below 40. Finally, when the SPL reaches 60-70 dB, the mid- to high-frequency features of the speech signal are completely masked, along with a substantial masking of low-frequency features. At this point, the overall GBO amplitude in the auditory cortex response decreases to the range of [0, 20], and the figure shows almost no visible GBO peaks.

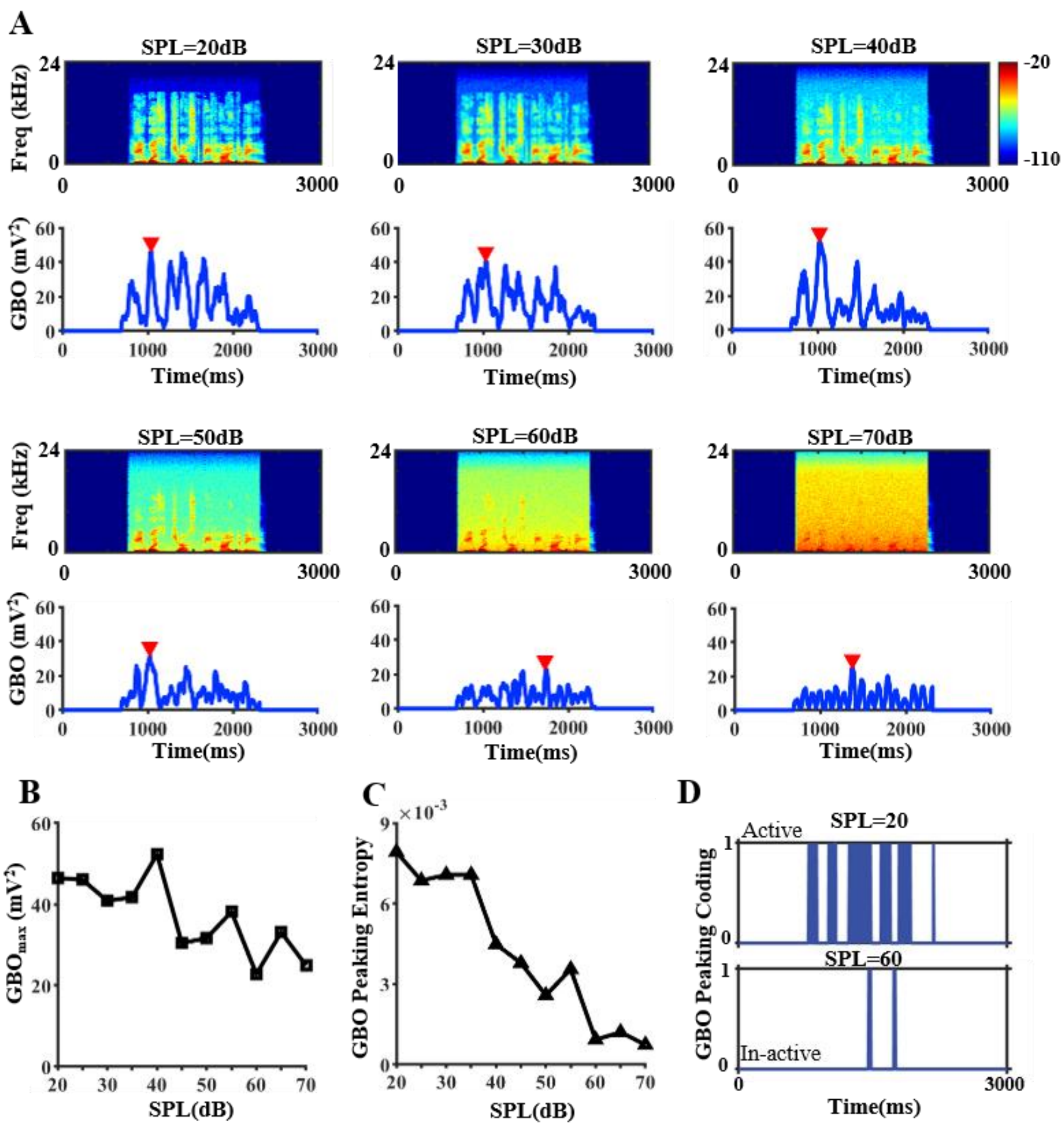

Fig 4 Auditory cortical neural network responses to speech signals mixed with different intensities of noise. A. Time-frequency characteristic maps of speech signals mixed with different intensity levels of noise and the GBO curves of the auditory cortex responses, with the peaks of the GBO curves highlighted by red downward triangles; B. the trend of the maximum of peaking values of the GBO curves in the auditory cortex responses under inputs of speech signals mixed with different intensities of noise; C. the trend of the peaking entropy in the GBO curves of the auditory cortex responses under inputs of speech signals mixed with different intensities of noise; D. the GBO peaking patterns as SPL is equal to 20 and 60, respectively.

In summary, when the noise SPL is below or equal to 40 dB, the masking effect of noise on the speech signal is weak, allowing the speech signal to retain many features, which corresponds to a higher level of brain activation, specifically reflected in a larger overall amplitude of the GBO curve, enabling individuals to perceive and encode the speech signal. When the noise SPL exceeds 40 dB, the masking effect intensifies, progressively masking the speech signal from high frequencies to mid- and low-frequency parts, which leads to lower brain activation levels, specifically indicated by a smaller overall amplitude of the GBO curve. At this point, individuals cannot clearly hear the speech, and their level of attention decreases accordingly. This result corroborates the experimental findings of Hahad et al. [45], which indicates that when environmental noise exceeds 40 dB, the brain's ability to recognize speech signals declines. Under such conditions, noise can affect sleep and rest, and when environmental noise exceeds 70 dB, it can interfere with conversations and influence work efficiency. Prolonged exposure to noise levels above 70 dB can also damage the auditory system. This indicates that the types and intensity parameters of noise selected in this study fall within a reasonable range for current cocktail party problem research.

Furthermore, by combining the dynamical GBO curves, we explored the evolution of the statistical maximum peaking amplitude and peaking entropy GBO curves as the noise intensity increased, as shown in Fig. 4B-D. Wherein, the efficient peaking point is marked only when the peaking value is over an appropriate threshold of 20, as shown for the GBO peaking patterns in Fig. 4D, based on which the GBO peaking entropy was obtained by computing the Shannon entropy algorithm of the GBO peaking coding information. It is evident that as the noise intensity parameter increases, the GBO peak values in the auditory cortical neural network exhibit an overall decreasing trend, dropping from the range of [40, 60] to the range of [20, 40]. Furthermore, when the noise intensity is between 40-70 dB, the decrease in GBO peak values is more pronounced. In addition, the increase in noise intensity also leads to a reduction in the peaking entropy in the GBO curve (Fig. 4C), in accompany with the degradation of the GBO peaking coding as shown in Fig. 4D. The above results imply that the peaks of the GBO curve and their spatiotemporal features can encode the features of speech signals. However, as noise intensity increases and the number of GBO curve peaks decreases, the corresponding features of the speech signal that can be encoded also diminish, leading to a decline in the ability of brain auditory system to perceive external speech signals.

In summary, the results indicate that as environmental noise intensity increases, its masking effect on speech signals intensifies, resulting in a decrease in the level of attention in the brain and a corresponding decline in the ability to perceive speech signals. Moreover, there exists a significant intensity threshold where noise exerts a marked effect, consistent with the experimental findings of Hahad et al. [45].

Furthermore, to study the oscillatory response patterns of the auditory cortical neural network under varying frequencies of auditory pure tone stimuli and noise stimuli, the pure tone stimuli with frequencies in the range of 200-1000 Hz and noise stimuli were separately introduced into the cortical auditory neural network model to analyze the auditory system response of discharge patterns to the stimuli. Moreover, for a comparison and verification for the model results in this paper with the experimental data, we also analyzed the GBO response patterns of the auditory cortical neural network based on the publicly available dataset with auditory EEG response to both the pure tone and noise stimuli [41], the results are shown in Fig. 5.

As shown in Fig. 5A, each pure tone selected was a single-frequency audio ranging from 200 Hz to 1000 Hz, while the noise was a specific decibel level of white noise. By analyzing the GBO responses of the cortical neural network, the results are presented in Fig. 5B and Fig. 5C. From Fig. 5B, it can be observed that as the frequency of the pure tone decreases, the amplitude of the GBO curve also decreases. In contrast, under white noise condition, the amplitude of the GBO curve is relatively low.

Fig. 5C shows the maximum values of each GBO curves for different stimuli. The black dots represent the maximal peak values of the GBO curve under the pure-tone stimulus, while the red dashed line indicates the maximal peak value of the GBO curve under white noise stimulation. It can be seen that as the frequency of the pure tone increases, the GBO peak values also show a gradual decreasing trend, while the GBO peak value under noise stimulation is below 20 and lower than the overall GBO peak values under the pure tone stimuli. This indicates that single-frequency of pure-tone input into auditory system can induce the cerebral cortex to generate gamma oscillations, which sustains the activation of auditory attention, while noise does not elicit the same response. Additionally, as the frequency of the pure tone decreases, the GBO peak value also decreases, suggesting that different frequency of pure tone can induce the brain to produce different activation degrees of gamma oscillations, implying that the frequency of pure tone affects the strength of attention. It has also been mentioned in previous works [46] that different intensity of sensory stimuli to human brain triggers the brain to generate varying degrees of gamma oscillations, further influencing the intensity of attention.

In Fig. 5D, the GBO curves of brain electrical responses under pure tone and noise stimuli were analyzed based on the publicly available dataset [41], it can be seen that as the auditory input of pure tones at 1000Hz are imposed to the brain, the EEG responses involve a similar peaking information of GBO curve. However, when the frequency of pure tone decreases to 500Hz, the GBO peaking amplitude obtains a similar decrease. Furthermore, when the auditory input is changed to white noise, that peaking information of the GBO in the EEG responses continues to go through a similar degradation as the modeling GBOs do (Fig. 5B). This validates the proposal of the modeling predictions that the GBO peaking information have the capacity to encode the auditory speech in a pure-tone (Fig. 5) and continuous-speech inputs.

Combining the results from Fig. 5B-D, it can be concluded that the brain can construct attention under pure tone stimuli, allowing individuals to focus on perceiving external information and improving the accuracy of information perception. In contrast, under noise stimulation, the gamma oscillations in the brain diminish, leading to a loss of attention and a diminished ability to accurately perceive external information. This further supports the experimental finding that gamma oscillations are key to constructing attention in the brain [47].

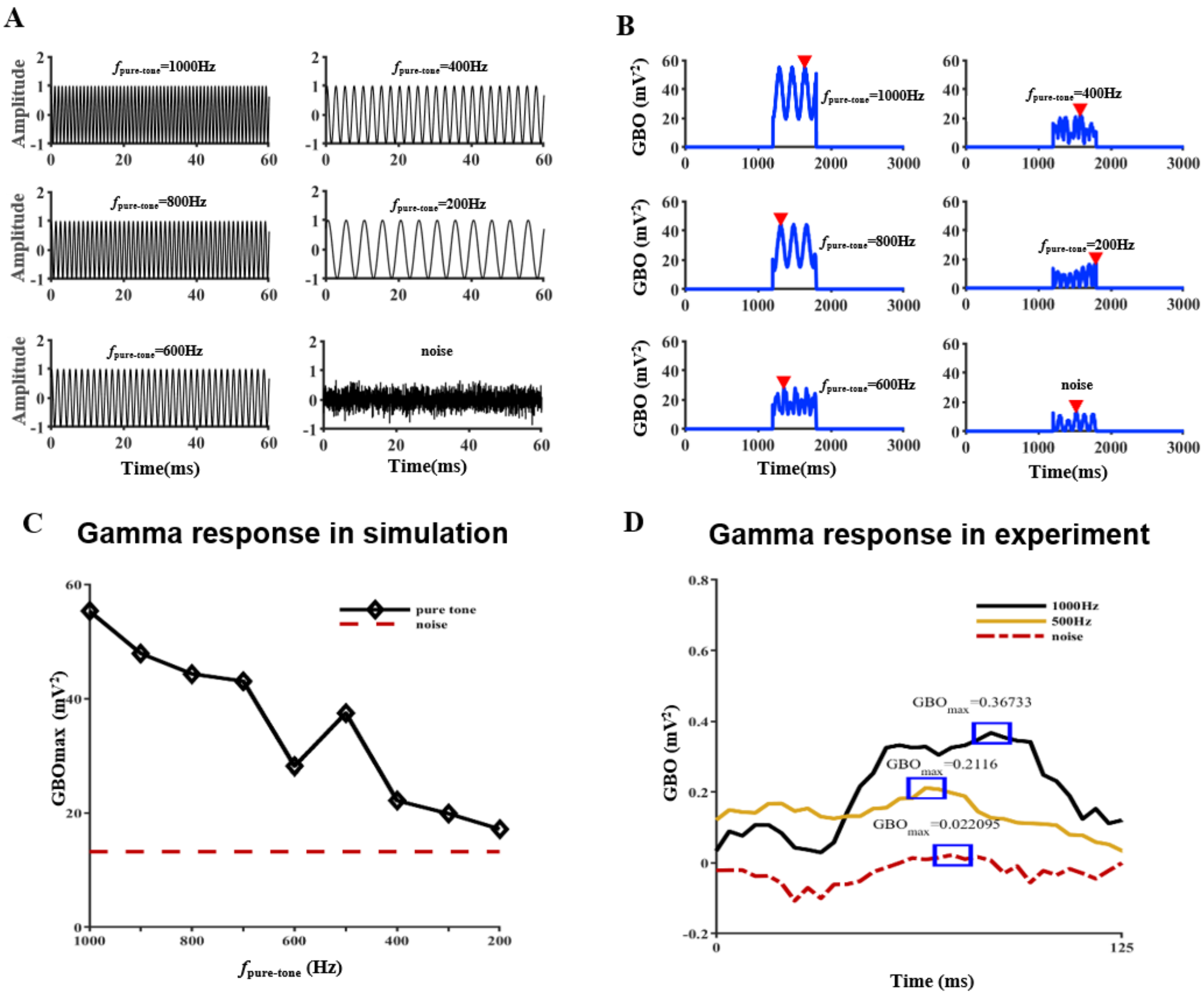

Fig. 5 Neural oscillatory response patterns of auditory cortical neural networks under pure -tone stimulation with frequency ranges from 200 to1000 Hz and noise stimulation. A. the depictions of 200-1000 Hz pure tone signals (increasing by 200 Hz) and noise signals; B. The GBO curves of the auditory cortex responses under 200-1000 Hz pure tone and noise stimulation, with the peak of the GBO curve highlighted by red downward triangles; C. A statistical chart of the maximum of peaking values of the GBO curves in the auditory cortex under 200-1000 Hz pure tone and noise stimulation; D. The GBO response to the1000Hz and 500Hz pure-tones and noise inputs based on experimental EEG data[45], with the maximum of the GBO curve marked by a blue rectangle.

### 3.3 Reverse control of synaptic excitatory-inhibitory balance on the attention-deficit GBO dynamics

To investigate the effects of varying excitatory-inhibitory levels on the construction process of auditory attention in the brain, we analyzed the effect of increase of the excitatory synaptic parameter $g_E$ on the discharge response of the auditory cortical neural network model. The analysis results are shown in Fig. 6. The increase of $g_E$ was discussed to simulate the condition when the excitatory synaptic connections increase, leading to the improvement of E-I ratio(the ratio between the mean values of EPSC(<EPSC>) and the IPSC(<IPSC>) in the auditory cortical neural network, as shown in Fig.6A. Fig. 6B describes the changes in the trend of E-I ratio, <EPSC> and <IPSC> in the model as $g_E$ increases. It can be seen that as $g_E$ increases, the E-I ratio in the model shows a gradual increasing trend, in company with a similar increase of <EPSC>. Meanwhile, the GBO curves involve significant improvement after $g_E$ increases from 0.1 to 0.6 (see Fig. 6C), and this GBO improvement can be manifested in the gradual increases of both the maximum and peaking coding entropy of the GBO curves as $g_E$ increases in the region [0.1, 0.6] with an increment of 0.5, as shown in Fig. 6D.

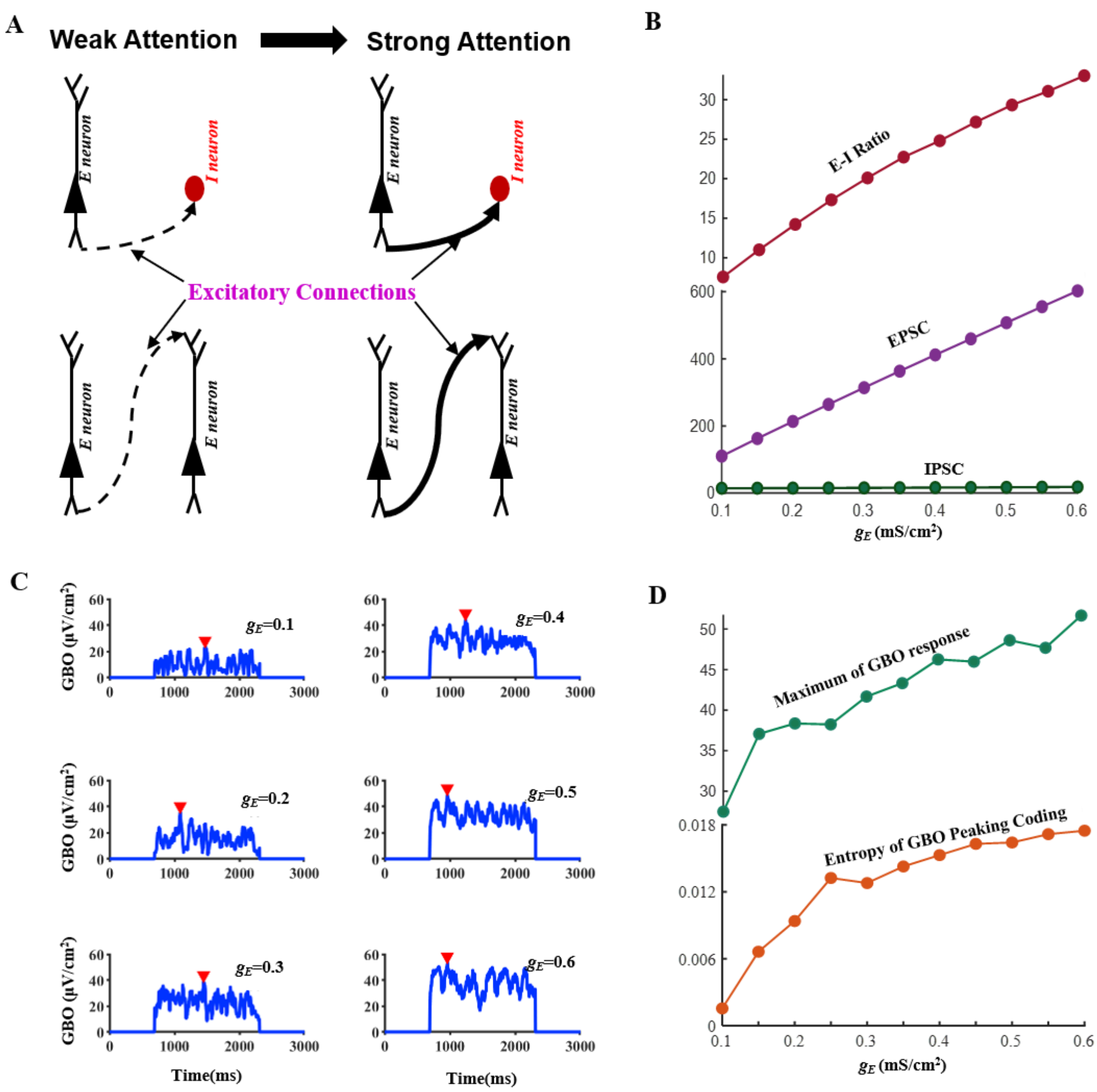

Fig. 6 The positive modulation effect of excitatory-inhibitory levels on brain attention construction A. The illustration of the effect of enhanced excitatory-inhibitory levels on attention construction. This includes two types of combinatory connections: pyramidal neuron-interneuron and pyramidal neuron-pyramidal neuron. The arrows representing synaptic connections on the left side of the figure are thinner, indicating weaker excitatory-inhibitory levels, while the arrows on the right side are thicker, indicating enhanced excitatory-inhibitory levels. The thick black arrow at the top of the figure describes the changes in attention before and after the enhancement of excitatory-inhibitory levels; B. The evolution trends of the E-I ratio (E-I ratio), excitatory EPSC, and IPSC in response to the increases of the excitatory-inhibitory parameter ($g_E$, excitatory synaptic conductance); C. GBO curves of auditory cortex responses in response to the different parameter of $g_E$ , with the maximum of the GBO curve highlighted by red downward triangles; D. the maximum values and the peaking coding entropy of each GBO curve in the auditory cortex as $g_E$ increases (increment of 0.5).

The above suggests that changes in the excitatory-inhibitory levels can not only influence the process of attention construction, but the coding capacity of the brain by modulating the strength and peaking coding of gamma oscillations. When the excitatory synaptic parameter increases, the excitatory connections also enhanced, as shown in Fig. 6A. At this point, the level of brain activation enhances, leading to increased excitation of brain neurons, which results in stronger gamma oscillations in the auditory cortical neural network, thereby concentrating attention and improving the precision of the brain's perception of external information. Increasing the brain's excitatory-inhibitory levels allows for the reconstruction of attention that had previously been disordered. This indicates that the brain can improve attention and enhance the accuracy of external information perception by adjusting excitatory-inhibitory levels [48].

Finally, in order to address the modulation effect of the excitation-inhibition balance parameter $g_E$, on the masking state under different conditions of background noise, this study examines the switching processes of the cortical neural network perceptual states in the parameter map of the noise intensity SPL and $g_E$. The states here are calibrated based on the value of GBO entropy, discussed in Fig. 6: first, the information entropy distribution of the GBO in the cortical neural network is calculated, and then a proper threshold is selected for normalizing the information entropy value into 0 and 1, where 0 represents the masking state, and 1 indicates the perceptual state. The selection of the noise intensity range is based on "Environmental noise in Europe, 2020" where urban noise is mostly between 50-70 dB, and commercial areas can exceed 75 dB [49]. Considering the intensity of noise in indoor scenarios, the range of noise intensity is selected to be between 20-80 dB. The variation range of the excitation-inhibition balance parameter $g_E$ is selected to be between 0.1 and 1.0.

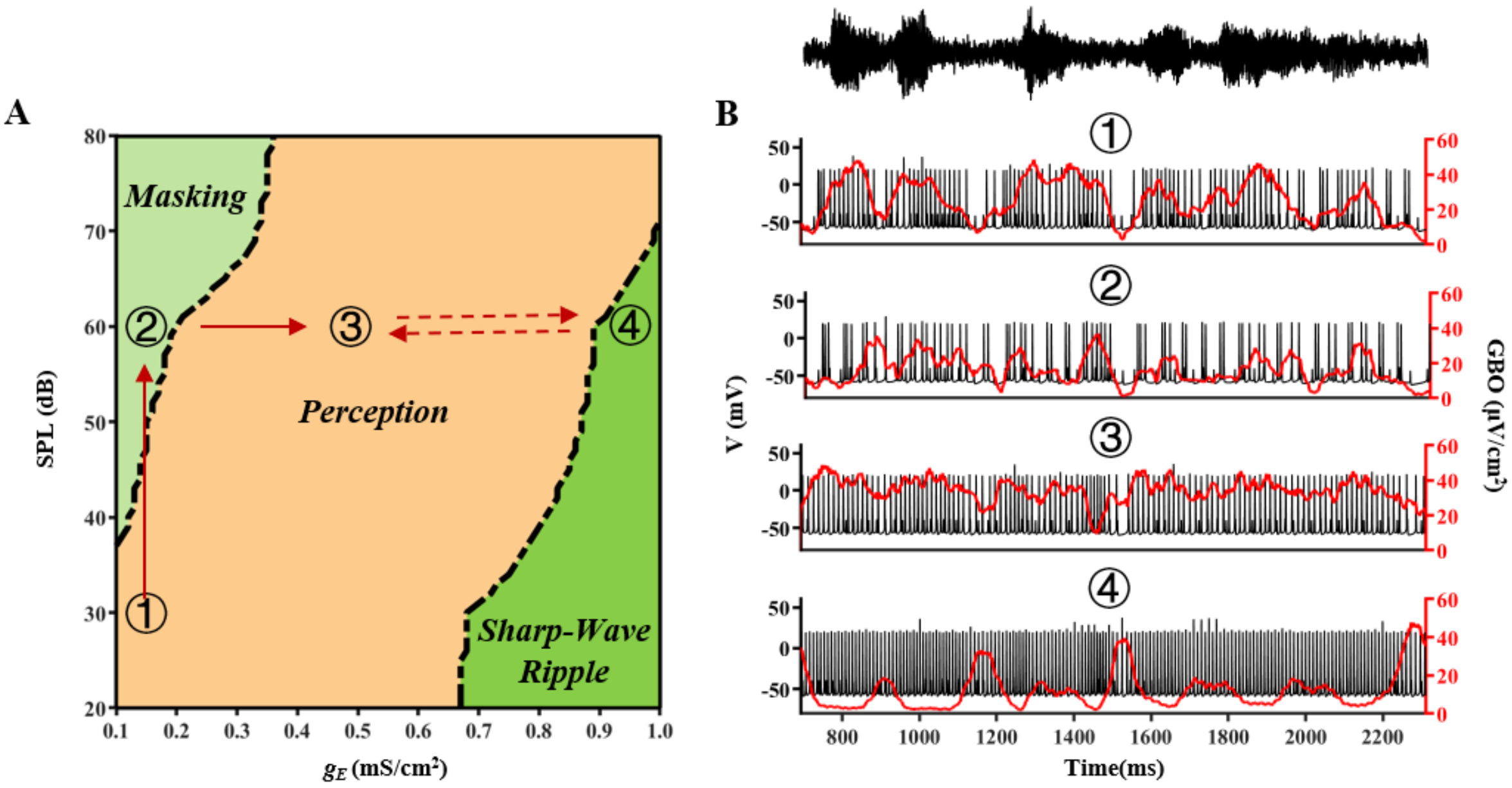

Fig. 7 Perceptual state switching of the cortical neural network. A. The light green and yellow areas in the left figure represent the masking state and the perceptual state, respectively, while the dark green area represents the *Sharp-Wave Ripple* (100 Hz - 200 Hz). The red arrows indicate the switching directions of the cortical perceptual states. B. The black lines in the figure represent neuronal firing, with unit of mV, while the red curve represents the dynamical curve of gamma oscillation(30-100Hz), with unit of mV/cm$^2$ .

As shown in Fig. 7, by adjusting the noise SPL and the value of $g_E$ to keep the cortical neural network in state ① *perception*, the noise intensity is relatively low, resulting in a good condition for the cortical neuronal network to encode the input of the target speech. From the neuron firing diagram of state ① in Fig.7.B, it can be observed that when the cortex perceives the target speech, the corresponding time peaks are denser, and the power of gamma oscillation(30-100Hz) is higher. At this point, the amount of information the cortex can perceive is large, demonstrating a bottom-up perceptual capability. As the noise intensity increases, the cortex gradually switches from state ① to state ② *masking*, where the noise intensity is higher, leading to a stronger masking effect on the target speech. From the neuronal firing of ② in Fig. 7.B, it can be seen that the dynamical power of the gamma oscillation corresponding to the target speech stimulus becomes lower than in state ①, indicating that the cortex is almost unable to perceive the target speech information and cannot complete bottom-up perception, thus being in a masking state. Furthermore, by increasing $g_E$, the cortex gradually switches from state ② to state ③. From the firing diagram of state ③, it can be seen that compared to state ②, the neurons are more excited, with denser peaks corresponding to the target speech stimulus period, and the gamma rhythm curve amplitude is increased back to the state ① level. In state ③, the cortex can perceive some of the target speech information and can complete bottom-up perception. At this point, the excitation-inhibition level in the neural network is elevated, improving the construction of cortical speech recognition attention, thereby altering the brain's perception intensity of the target speech. Based on state ③, if the value of $g_E$ is continued to be increased, the excitation-inhibition balance level in the cortex is further enhanced, gradually switching from state ③ to state ④ *Sharp−Wave Ripple*. If the process between state ③ and state ④ can be adjustably switched by the modulation of brain excitation-inhibition balance level, then the cortical response to *Sharp-Wave Ripple* could represent processes of higher cognitive activities of attention selection, memory, recall after the brain percept the bottom-up information [50,51]. From the firing diagram of state ④, it can be seen that compared to state ③, the peaks corresponding to the target speech stimulus period are denser, and the overall firing frequency of the neurons is higher 100−200Hz. Therefore, the amplitude of the gamma oscillation (30−100Hz) decreases, allowing the cortex to perceive more information. Under this hypothesis, the cortex can not only complete the bottom-up perception process but may also encode the input stimuli and engage in working memory processes to achieve complete top-down modulation [52,53]. However, if the process from state ③ to state ④ cannot be reversable or adjustable, then the abnormal excitation-inhibition level leading to the cortical response of *Sharp-Wave Ripple* could be a characteristic marker of diseases such as epilepsy [54].

**4 Discussion：**

The construction of auditory perceptual attention is crucial for understanding speech processing in noisy environments and solving the cocktail party problem. Previous work has used a black - box model to quantify the dynamic relationship between input speech signals and neural responses, focusing on the average discharge rate of neuronal populations [55]. However, the lack of model interpretability has made it difficult to understand how neural responses encode auditory inputs in detail. This paper addresses this gap by employing an interpretable cortical neural network model to simulate the auditory loop. Through analyzing the model's dynamic neural oscillations, we found that under continuous speech and pure tone speech inputs, neural population responses exhibit specific gamma oscillation amplitude encodings. These findings are consistent with previous experiments that observed similar gamma oscillation amplitude encodings in responses to transient auditory stimuli [32,56,57]. Our study not only demonstrates gamma oscillation amplitude encoding for pure tones but also reveals that continuous speech input significantly increases the dynamic gamma oscillation peak amplitude and times, enhancing the gamma oscillation encoding capability of entropy for speech inputs (Fig. 4C). This suggests that gamma oscillation peaking patterns may represent potential encoding forms used by the auditory cortex for processing continuous natural language [57].

Our previous research highlighted the key role of cortical gamma oscillations in building attention [31]. The simulation results presented in this paper further confirm that noise in the brain auditory system reduces the auditory cortex's gamma oscillation response encoding capability, as evidenced by gamma oscillation amplitude attenuation (Fig. 4, Fig. 5). Additionally, our analysis of publicly available experimental data reveals that the amplitude attenuation of gamma oscillations induced by noise aligns closely with our model's predictions (Fig. 5D), reinforcing the importance of gamma oscillations in constructing attention for actual speech inputs.

We also utilized nonlinear system stability analysis to explore the parameter conditions under which attention - related gamma oscillations occur. The results indicate that as noise masking decreases (i.e., attention levels increase), gamma oscillations become easier to generate, with a significant increase in the excitatory currents of EPSCs to individual neurons. This further confirms the key role of chemical excitatory synapses in the generation of synchronized gamma oscillations [58,59]. Nevertheless, considering the complex synchronization characteristics of neural networks [60], investigating the underlying synchronization transition dynamics remains a necessary direction for future research.

In the field of artificial neural networks, attention mechanisms have been shown to improve information recognition accuracy [61]. Similarly, our analysis indicates that gamma oscillations in the auditory cortex aid in constructing auditory attention, with the strength of gamma oscillations decreasing under reduced attention (masking) conditions. This underscores the key role of cortical gamma oscillations related to attention mechanisms in the brain auditory system's natural speech information recognition. These findings are consistent with our previous experimental studies [31] and suggest that in artificial neural networks, attention can be adaptively designed and optimized based on the average activation values of the neural network.

Finally, we examined how varying excitatory - inhibitory balance levels affect attention - dependent gamma oscillations and the reverse modulation pathways for attention masking. The results reveal that increasing the brain's excitatory - inhibitory balance level enhances speech recognition attention construction. Previous studies have shown that altering the excitatory - inhibitory balance in the cerebral cortex can modulate the dynamics of neuronal assemblies in cortical networks [48]. Scientists have also used this principle in combination with optogenetic stimulation to regulate abnormal brain states [62]. Based on our findings, increasing cortical excitatory - inhibitory levels through optogenetic stimulation could improve auditory attention construction, offering a potential therapeutic approach for auditory attention disorders such as hearing loss and auditory hallucinations. Moreover, enhancing the excitatory - inhibitory balance level can alter the brain's perception intensity of target speech, indicating an intrinsic brain mechanism for controlling selective attention in complex environments [63]. Recent experiments have also highlighted the potential involvement of the brain's dopamine system in specific speech recognition within complex auditory environments [64,65]. Future research combining dopamine and other intrinsic brain circuit mechanisms will be crucial for understanding the cocktail party problem in auditory language processing.

**Acknowledgement**

We thank the Delorme and his colleagues for provide publicly available EEG data for us accessible to validate the simulation results. This work was supported by the National Natural Science Foundation of China (Grant Nos. 12572068, 12002251).

**Data Availability Statement**

A portion or all data, models, and code generated or used during the study are available from the corresponding author upon request.

**Conflict of Interest**

The authors declare that they have no conflict of interest.

**Appendix:**

In the Materials and Method section of each excitatory/inhibitory neuron, we have adopted the classical models to describe their dynamics.

The two-compartment model for describing dynamics of the pyramidal soma is described by the following ordinary differential equations by referring the literature [38]:

$$C_m \frac{dV_E}{dt} = f^E(V_E, m, n, h) + \frac{g_c}{p}\left(V_d - V_E\right) \tag{A-1}$$

After expanding the ionic current function in the equation $f^E(V_E, m, n, h)$, the complete equation is as follows:

$$C_{m,E} \frac{dV_E}{dt} = g_L\left(V_L - V_E\right) + g_K n^4\left(V_K - V_E\right) + g_{Na} m_\infty^3 h\left(V_{Na} - V_E\right) + \frac{g_c}{p}\left(V_d - V_E\right) \tag{A-2}$$

Where $g_L\left(V_L - V_E\right)$ represents the leakage current, $g_L$ is the leakage conductance, which is an ion channel on the neuron membrane that allows ions to pass through, resulting in the leakage current. $g_K n^4\left(V_K - V_E\right)$ represents the potassium ionic current, where $g_K$ is the potassium ionic conductance. $g_{Na} m_\infty^3 h\left(V_{Na} - V_E\right)$ represents the sodium ionic current, where $g_{Na}$ is the sodium ionic conductance. $m$、$n$、$h$ are gating variables that describe the opening and closing states of the ion channels.

The pyramidal dendritic neuron is modelled as follows [38]:

$$C_{m,E} \frac{dV_{Ed}}{dt} = f^{Ed}(V_{Ed}, s, \left[Ca^{2+}\right]_n) + \frac{g_c}{\left(1-p\right)}\left(V_E - V_{Ed}\right) + I_{syn}^{Ed} + I_{thalamus}^{Ed} \tag{A-3}$$

After expanding the ionic current function in the equation $f^{Ed}(V_{Ed}, s, \left[Ca^{2+}\right]_n)$, the complete equation is as follows:

$$C_{m,E}\frac{dV_{Ed}}{dt}=g_{Ca}s\left(V_{Ca}-V_{Ed}\right)+g_L\left(V_L-V_{Ed}\right)+g_{AHP}\left[\left[Ca^{2+}\right]/\left(\left[Ca^{2+}\right]+K_D\right)\right]\left(V_K-V_{Ed}\right)+\frac{g_c}{(1-p)}\left(V_s-V_{Ed}\right)+I_{syn}^{Ed}+I_{thalamus}^{Ed} \tag{A-4}$$

Where $g_{Ca}s\left(V_{Ca}-V_{Ed}\right)$ represents the leakage current, $g_{Ca}$ is the leakage conductance, which is an ion channel on the neuron membrane that allows ions to pass through, resulting in the leakage current. $s$ is a gating variable that describes the opening and closing states of the ion channels. $g_L\left(V_L-V_{Ed}\right)$ represents the leakage current, where $g_L$ is the leakage conductance. $g_{AHP}\left[\left[Ca^{2+}\right]/\left(\left[Ca^{2+}\right]+K_D\right)\right]\left(V_K-V_{Ed}\right)$ represents the calcium-activated potassium current. Other variables are explained in Eq. (7).

In the above equations, the gating variables $m$、$n$、$h$、$s$ are described by the following equations:

$$m=\alpha_m/\left(\alpha_m+\beta_m\right) \tag{A-5}$$

$$\alpha_m=-0.1\left(V_s+33\right)/\left\{\exp\left[-0.1\left(V_s+33\right)\right]-1\right\} \tag{A-6}$$

$$\beta_m=4\exp\left[-\left(V_s+58\right)/12\right] \tag{A-7}$$

$$\frac{dh}{dt}=\theta_h\left[\alpha_h\left(1-h\right)-\beta_h h\right] \tag{A-8}$$

$$\alpha_h=0.07\exp\left[-\left(V_s+50\right)/10\right] \tag{A-9}$$

$$\beta_h=1/\left\{\exp\left[-0.1\left(V_s+20\right)\right]+1\right\} \tag{A-10}$$

$$\frac{dn}{dt}=\theta_n\left[\alpha_n\left(1-n\right)-\beta_n n\right] \tag{A-11}$$

$$\alpha_n=-0.01\left(V_s+34\right)/\left\{\exp\left[-0.1\left(V_s+34\right)\right]-1\right\} \tag{A-12}$$

$$\beta_n=0.125\exp\left[-\left(V_s+44\right)/25\right] \tag{A-13}$$

$$s=1/(1+\exp(-(V_d+20)/9)) \tag{A-14}$$

The intracellular $Ca^{2+}$ concentration is controlled by the leakage integral:

$$\frac{d\left[Ca^{2+}\right]}{dt}=-\alpha I_{Ca}-\left[Ca^{2+}\right]/\tau_{Ca} \tag{A-15}$$

Where $\alpha$ is proportional to *S/V* (membrane area/volume directly beneath the membrane). In this paper, $\alpha$ =0.002, with units of μM(ms·μA)$^{-1}$cm$^2$, and the amount of $Ca^{2+}$ influx for each spike is 200 nM. The various squeezing and buffering mechanisms of $Ca^{2+}$ ions in channels are collectively described by a single exponential decay process with a time constant $\tau_{Ca}$=80ms. The corresponding parameters in the above equations are presented in Table 1.

Table 1 Parameters for pyramidal neuron model [38]

| Parameters | Values | Parameters | Values |
|---|---|---|---|
| $C_{E,m}$ | 1.0μF/cm$^2$ | $V_L$ | -65mV |
| $g_L$ | 0.1mS/cm$^2$ | $V_{Na}$ | 55mV |
| $g_{Na}$ | 45mS/cm$^2$ | $V_K$ | -80mV |
| $g_K$ | 18mS/cm$^2$ | $V_{ca}$ | 120mV |
| $g_c$ | 2.0mS/cm$^2$ | $\theta_h$ | 4 |
| $p$ | 0.5 | $\theta_n$ | 4 |
| $g_{Ca}$ | 1mS/cm$^2$ | $\alpha$ | 0.002 |
| $g_{AHP}$ | 0.5mS/cm$^2$ | $\tau_{Ca}$ | 80mS |
| $K_D$ | 30 | | |

The fast-spiking interneuron model is described by the following ordinary differential equations:

$$C_{m,I}\frac{dV_I}{dt}=f^I(V_I,m,n,h) \tag{A-16}$$

After introducing the expansion form of the ionic current function $f^I(V_I,m,n,h)$ based on the literature [33], the complete equation for describing the neural dynamics of single interneuron is shown as follows:

$$C_{m,I}\frac{dV_I}{dt}=g_{Na}m^3h\left(V_{Na}-V_I\right)+g_Kn^4\left(V_K-V_I\right)+g_L\left(V_L-V_I\right) \tag{A-17}$$

Where $g_L\left(V_L-V_I\right)$ represents the leakage current, $g_L$ is the leakage conductance, which is an ion channel on the neuron membrane that allows ions to pass through, resulting in leakage current. $g_Kn^4\left(V_K-V_I\right)$ represents the potassium ionic current, where $g_K$ is the potassium ionic conductance. $g_{Na}m^3h\left(V_{Na}-V_I\right)$ represents the sodium ionic current, where $g_{Na}$ is the sodium ionic conductance. The variables $m$、$n$、$h$ are gating variables that describe the opening and closing states of the ion channels.

In the above equations, the gating variables $m$、$n$、$h$ are described by the following equations:

$$m=\alpha_m/\left(\alpha_m+\beta_m\right) \tag{A-18}$$

$$\alpha_m=-0.1\left(V+35\right)/\left\{\exp\left[-0.1\left(V+35\right)\right]-1\right\} \tag{A-19}$$

$$\beta_m=4\exp\left[-\left(V+60\right)/18\right] \tag{A-20}$$

$$\frac{dh}{dt}=\theta_h^I\left[\alpha_h\left(1-h\right)-\beta_hh\right] \tag{A-21}$$

$$\alpha_h=0.07\exp\left[-\left(V+58\right)/20\right] \tag{A-22}$$

$$\beta_h = 1/\left\{\exp\left[-0.1(V+28)\right]+1\right\} \tag{A-23}$$

$$\frac{dn}{dt} = \theta_n^I\left[\alpha_n(1-n)-\beta_n n\right] \tag{A-24}$$

$$\alpha_n = -0.01(V+34)/\left\{\exp\left[-0.1(V+34)\right]-1\right\} \tag{A-25}$$

$$\beta_n = 0.125\exp\left[-(V+44)/80\right] \tag{A-26}$$

The corresponding parameters for interneurons in the above equations are presented in Table 2.

Table 2 Parameters for the fast-spiking interneuron model [33]

| Parameters | Values | Parameters | Values |
|---|---|---|---|
| $C_{I,m}$ | 1.0μF/cm$^2$ | $V_{Na}$ | 55mV |
| $g_L$ | 0.1mS/cm$^2$ | $V_K$ | -90mV |
| $g_{Na}$ | 35mS/cm$^2$ | $\theta^I_h$ | 5 |
| $g_K$ | 9mS/cm$^2$ | $\theta^I_n$ | 5 |
| $V_L$ | -65mV | | |